\documentclass[10pt,conference]{IEEEtran}
\IEEEoverridecommandlockouts

\usepackage{cite}
\usepackage[utf8]{inputenc}
\usepackage{amsmath,amssymb,amsfonts}
\usepackage{algorithm}
\usepackage{algorithmicx}
\usepackage{algpseudocode}
\usepackage{graphicx}
\usepackage{textcomp}
\usepackage{xcolor}
\usepackage{balance}
\usepackage{makecell}
\usepackage{pifont}
\usepackage{url}
\usepackage{booktabs}
\usepackage{tabularx}
\usepackage{enumitem}
\usepackage{hyperref} 
\usepackage{subcaption}
\setlist{noitemsep,topsep=1pt}
\AtBeginDocument{%
\setlength{\abovedisplayskip}{5pt plus 1pt minus 1pt}%
\setlength{\belowdisplayskip}{5pt plus 1pt minus 1pt}%
\setlength{\abovedisplayshortskip}{4pt plus 1pt minus 1pt}%
\setlength{\belowdisplayshortskip}{4pt plus 1pt minus 1pt}%
}
\DeclareRobustCommand{\myurl}[1]{\url{#1}}

\newcommand{\sem}{\mathrm{S}}
\newcommand{\conv}{\mathrm{C}}
\newcommand{\pTwoP}{\mathrm{P2P}}
\newcommand{\Uu}{\mathrm{Uu}}
\newcommand{\SL}{\mathrm{SL}}
\newcommand{\pay}{\mathrm{pay}}
\newcommand{\oh}{\mathrm{oh}}
\newcommand{\sync}{\mathrm{sync}}
\newcommand{\ctrl}{\mathrm{ctrl}}
\newcommand{\fb}{\mathrm{fb}}
\newcommand{\cmp}{\mathrm{cmp}}
\newcommand{\rs}{\mathrm{rs}}

\newcommand{\meta}{\mathrm{meta}}
\newcommand{\tx}{\mathrm{tx}}
\newcommand{\rx}{\mathrm{rx}}

\def\BibTeX{{\rm B\kern-.05em{\sc i\kern-.025em b}\kern-.08em T\kern-.1667em\lower.7ex\hbox{E}\kern-.125emX}}

\newcommand{\todo}[1]{\textsf{\textbf{\textcolor{Orange}{[#1]}}}}

\begin{document}
\title{
The Price of Meaning: Quantifying Semantic Communication Overheads in Practice
}

\author{\IEEEauthorblockN{
Xinyi Lin, Peizheng Li, Adnan Aijaz
}\\ 
\IEEEauthorblockA{
Bristol Research and Innovation Laboratory, Toshiba Europe Ltd., U.K.\\
Email: 
 {\{xinyi.lin, peizheng.li, adnan.aijaz\}@toshiba-bril.com}
}
}

\maketitle


\begin{abstract}
Semantic communication (SemCom) promises to reduce transmitted payloads by conveying task-relevant meaning instead of raw bits. However, practical SemCom also incurs semantic metadata, control signaling, feedback, model or knowledge-base synchronization, and neural computation costs, which may offset semantic compression gains. This paper develops an overhead-aware analytical framework for quantifying the spectral-resource and energy costs of SemCom under equal task utility. The framework covers point-to-point transmission, user equipment (UE)-to-next-generation NodeB (gNB) uplink, and UE-to-UE communication under a single gNB, and derives closed-form break-even conditions with respect to payload size, semantic compression factor, model reuse, protocol overhead, and computation energy. Simulation results show that SemCom becomes spectrally beneficial only for sufficiently large payloads, while energy gains require larger payloads due to processing and synchronization overheads. The results also show that multi-user downlink is particularly favorable, as shared semantic overheads can be amortized across multiple UEs. These findings provide design guidance for realistic SemCom evaluation and standardization-oriented deployment.
\end{abstract}


\section{Introduction}
\IEEEPARstart{S}{emantic} communication (SemCom) revisits the classical communication objective by emphasizing the transfer of meaning, task utility, or receiver action rather than bit-exact reconstruction. Early information theory deliberately separated the technical transmission problem from semantic and effectiveness problems \cite{Shannon1948,Weaver1949}; modern SemCom reopens this separation by using learned representations, knowledge bases, and task-oriented metrics. Neural SemCom systems such as DeepSC have shown that semantic representations can be robust at low signal-to-noise ratio (SNR) and may reduce the amount of payload that must be transmitted \cite{Xie2021DeepSC}. Surveys and tutorials further highlight SemCom as a candidate enabler for 6G intelligence, sensing, and edge-native applications \cite{Lu2024Survey}.

Despite these advances, a key practical question remains unresolved: \emph{what is the price of meaning?} A semantic payload is not transmitted in isolation. Real systems must identify the semantic task, model version, embedding format, context, confidence level, and reliability target. They may also require semantic feedback, channel feedback, retransmission triggers, and model or knowledge-base synchronization. In cellular deployments, these semantic-specific costs coexist with unavoidable 5G New Radio (NR) control and reference-signal overheads, including scheduling, uplink grants, physical uplink control channel (PUCCH), physical downlink control channel (PDCCH), demodulation reference signals (DMRS), hybrid automatic repeat request (HARQ), and radio resource control (RRC) procedures \cite{TS38212,TS38213,TS38331}. Thus, reducing the application payload does not automatically imply reducing the total radio or energy cost.

This paper develops a compact analytical model for overhead-aware SemCom evaluation. Rather than asking whether semantic representations are shorter than conventional payloads, we ask whether SemCom reduces the \emph{total} cost needed to achieve a target utility. We consider both communication-resource and energy cost, and we instantiate the model in three deployment scenarios: point-to-point communication, UE-to-gNB uplink communication, and UE-to-UE communication within a single gNB coverage area. The resulting expressions expose the conditions under which SemCom is beneficial and the regimes where overhead dominates.

\section{Related Work and Contributions}
Existing SemCom studies can be grouped into four categories. First, foundational and tutorial works define the SemCom vision, semantic metrics, knowledge support, and implementation choices \cite{Lu2024Survey}. Second, neural end-to-end systems, including text, speech, and image SemCom, demonstrate robustness and payload reduction relative to conventional bit-oriented baselines \cite{Xie2021DeepSC}. Third, resource-allocation studies introduce semantic spectral efficiency and optimize channel or semantic-symbol allocation 
\cite{11274758}. Fourth, recent energy-aware works account for the cost of semantic extraction, inference, or communication--computation tradeoffs. For example, rate-splitting-based SemCom minimizes communication and computation energy \cite{Yang2023RS}, green transformer selection benchmarks semantic loss against CPU/GPU energy \cite{Mukherjee2023Green}, and probabilistic SemCom over space-air-ground integrated networks models the tradeoff between semantic compression and computation energy \cite{Zhao2024PSC}. Feedback-aware SemCom has also been studied for reliability, where adaptive channel feedback is allocated based on predicted semantic distortion \cite{Zhang2023SCAN}.

These works establish that SemCom can improve task performance, spectral efficiency, or energy efficiency under suitable assumptions. However, most evaluations focus on semantic payloads, semantic symbols, transmit energy, or inference energy. They usually do not jointly quantify the practical overhead stack required for deployment: protocol control signaling, semantic metadata, reliability feedback, model or knowledge-base synchronization, and UE-side computation. This gap is especially important for short packets, where fixed control and synchronization overhead can exceed the semantic payload itself.

This paper makes the following contributions:
\begin{itemize}
    \item We develop an overhead-aware SemCom cost framework that jointly accounts for semantic payload compression, metadata, control signaling, feedback, reference signals, model/knowledge synchronization, and endpoint computation.
    \item We instantiate the framework for practical wireless deployments, including point-to-point links, NR  UE--gNB uplink radio interface, gNB-to-multi-UE downlink, sidelink UE-to-UE communication, and network-routed UE-to-UE communication under one gNB.
    \item We derive closed-form spectral-resource and energy break-even conditions that expose the roles of payload size, semantic compression factor, model reuse interval, protocol overhead, and neural computation energy.
    \item We provide numerical results showing when SemCom is beneficial in practice: spectral gains appear after fixed overheads are amortized, energy gains require stricter conditions, multi-user downlink offers strong overhead sharing, and large model reuse with strong compression is critical.
\end{itemize}
\vspace{-1mm}
\section{System Model}
\vspace{-1mm}
\subsection{Utility-Constrained Communication}
\vspace{-1mm}
Consider a source message associated with a task utility target $U_0$. A conventional transmitter sends a source-coded payload of length $L$ bits. A semantic transmitter maps the source to a task-relevant representation using encoder $f_{\theta}$ and decoder $g_{\phi}$ with shared context or knowledge base $\mathcal{K}$, where $\theta$ and $\phi$ are the encoder and decoder model parameters. For a target utility $U_0$, define the semantic compression factor
\begin{equation}
    0 < \rho(U_0) \leq 1,
\end{equation}
so that the semantic payload is
\begin{equation}
    B_{\sem,\pay} = \rho(U_0)L,
\end{equation}
whereas the conventional payload is $B_{\conv,\pay}=L$. The parameter $\rho(U_0)$ captures the fact that stricter utility, distortion, or task-accuracy requirements reduce semantic compressibility.

Let $s \in \{\pTwoP,\Uu,\SL\}$ index the deployment scenario, corresponding to point-to-point (P2P), NR Uu, and sidelink (SL) communication, respectively. Let $x\in\{\conv,\sem\}$ index the communication mode, where $\conv$ and $\sem$ denote conventional and semantic communication. The evaluation target is
\begin{equation}
    \eta_{\mathrm{eff},x}^{(s)} = \frac{U_x}{A_x^{(s)}}, \qquad
    \xi_{\mathrm{eff},x}^{(s)} = \frac{U_x}{E_x^{(s)}},
\end{equation}
where $U_x$ is the achieved utility, $A_x^{(s)}$ is the total spectral-resource cost, and $E_x^{(s)}$ is the total energy cost. A fair comparison is made at equal utility, i.e., $U_\sem \ge U_0$ and $U_\conv \ge U_0$.

\subsection{Generic Overhead Taxonomy}
The conventional total information-equivalent burden is
\begin{equation}
    B_{\conv}^{\mathrm{tot}} = L + B_{\conv,\ctrl}^{(s)} + B_{\conv,\fb}^{(s)} + B_{\conv,\mathrm{hdr}}^{(s)} ,
    \label{eq:convbits}
\end{equation}
where $B_{\conv,\ctrl}^{(s)}$, $B_{\conv,\fb}^{(s)}$ and $B_{\conv,\mathrm{hdr}}^{(s)}$ denote the control-channel, feedback, and protocol header burden, respectively. 
For SemCom, the burden is
\begin{equation}
    B_{\sem}^{\mathrm{tot}} = \rho L + B_{\meta}^{(s)} + B_{\sem,\ctrl}^{(s)} + B_{\sem,\fb}^{(s)} + \frac{B_{\sync}^{(s)}}{N},
    \label{eq:sembits}
\end{equation}
where $B_{\meta}^{(s)}$ includes model identifier, task identifier, embedding format, context index, confidence information, semantic quality-of-service (QoS) fields, and tokenizer or codebook version. The term $B_{\sync}^{(s)}$ is the traffic required to distribute or update the semantic model, codebook, probability graph, or knowledge base, and $N$ is the number of payloads over which this cost is amortized. Thus, small $N$ represents dynamic tasks or frequent model updates, while large $N$ represents stable shared context.

\subsection{Spectral-Resource Cost}
Let $\mathcal{H}_s$ denote the set of wireless hops in scenario $s$. The total spectral-resource cost is modeled as
\begin{align}
A_x^{(s)} = &\sum_{\ell\in\mathcal{H}_s}
\frac{B_{x,\ell}^{\pay}+B_{x,\ell}^{\meta}}{\eta_{x,\ell}}
+ A_{x,\ctrl}^{(s)} + A_{x,\fb}^{(s)}\nonumber \\ 
& + A_{x,\rs}^{(s)} + \frac{A_{x,\sync}^{(s)}}{N},
\label{eq:spectralgeneric}
\end{align}
where $\eta_{x,\ell}$ is the payload spectral efficiency on hop $\ell$, $B_{x,\ell}^{\pay}$ and $B_{x,\ell}^{\meta}$ are the payload and metadata bits transmitted over hop $\ell$, $A_{x,\ctrl}^{(s)}$ is the control-channel resource cost, $A_{x,\fb}^{(s)}$ is the feedback resource cost, $A_{x,\rs}^{(s)}$ accounts for pilots, DMRS, or other reference signals, and $A_{x,\sync}^{(s)}$ is the spectral cost of model or knowledge-base synchronization. Eq. \eqref{eq:spectralgeneric} intentionally separates payload compression from fixed and semi-fixed protocol costs.

\subsection{Energy Cost}
The total energy cost is
\begin{equation}
E_x^{(s)} = E_{x,\mathrm{air}}^{(s)} + E_{x,\ctrl}^{(s)} + E_{x,\fb}^{(s)} + E_{x,\cmp}^{(s)} + \frac{E_{x,\sync}^{(s)}}{N},
\label{eq:energygeneric}
\end{equation}
where $E_{x,\mathrm{air}}^{(s)}$, $E_{x,\ctrl}^{(s)}$, $E_{x,\fb}^{(s)}$, $E_{x,\cmp}^{(s)}$, and $E_{x,\sync}^{(s)}$ denote the air-interface, control signaling, feedback, computation, and synchronization energy, respectively, with the synchronization cost amortized over $N$ payloads. 
The air-interface energy is
\begin{equation}
E_{x,\mathrm{air}}^{(s)} = \sum_{\ell\in\mathcal{H}_s} \left(P_{\ell}^{\tx}T_{x,\ell}^{\tx}+P_{\ell}^{\rx}T_{x,\ell}^{\rx}\right),
\end{equation}
where
\begin{equation}
T_{x,\ell}^{\tx} \approx T_{x,\ell}^{\rx} = \frac{B_{x,\ell}^{\pay}+B_{x,\ell}^{\meta}}{W_{\ell}\eta_{x,\ell}} .
\end{equation}
In the air-interface energy model, $P_{\ell}^{\tx}$ and $P_{\ell}^{\rx}$ are the transmit and receive power on hop $\ell$, $T_{x,\ell}^{\tx}$ and $T_{x,\ell}^{\rx}$ are the corresponding transmission and reception duration, $B_{x,\ell}^{\pay}$ and $B_{x,\ell}^{\meta}$ are the payload and metadata bits, $W_{\ell}$ is the transmission bandwidth, and $\eta_{x,\ell}$ is the spectral efficiency.
For neural SemCom, computation energy is
\begin{equation}
E_{\sem,\cmp}^{(s)} = e_{\mathrm{MAC}}(C_{\mathrm{enc}}+C_{\mathrm{dec}}) + e_{\mathrm{mem}}M_{\mathrm{acc}} + E_{\mathrm{post}},
\label{eq:compenergy}
\end{equation}
where $C_{\mathrm{enc}}$ and $C_{\mathrm{dec}}$ are encoder and decoder operation counts, $e_{\mathrm{MAC}}$ is the energy per multiply--accumulate operation, $M_{\mathrm{acc}}$ is the memory-access count, $e_{\mathrm{mem}}$ is the energy consumed per memory-access, and $E_{\mathrm{post}}$ covers task-specific post-processing.

In the numerical evaluation, $E_{x,\cmp}^{(s)}$ denotes endpoint computation needed to meet $U_0$. For conventional communication it includes source coding, decoding, and post-processing; for SemCom it also includes semantic encoding/decoding, inference, memory movement, and semantic decision processing. This separation avoids counting transmit-bit reduction as an energy gain unless computation and synchronization energy are also amortized, especially for battery-limited UEs.

\subsection{Deployment Scenarios}
We consider three scenarios.

\subsubsection{Scenario A: Point-to-Point}
A transmitter communicates directly with a receiver over one hop, i.e., $\mathcal{H}_{\pTwoP}=\{1\}$, which isolates semantic overhead from cellular scheduling overhead. The semantic-specific costs are metadata, semantic feedback, alignment messages, and amortized model synchronization.

\subsubsection{Scenario B: UE-to-gNB over Uu}
A UE transmits to a serving gNB through the NR Uu uplink. In addition to payload transmission on the physical uplink shared channel (PUSCH), the UE pays for scheduling requests, grants, buffer status reporting, PDCCH monitoring, PUCCH feedback, DMRS, HARQ, and higher-layer headers. SemCom further adds task/model metadata, semantic-quality reports, and UE-side encoding energy.

\subsubsection{Scenario C: UE-to-UE within a Single gNB}
Two UEs communicate under one gNB. The primary case considered is gNB-controlled NR sidelink, where the useful payload is carried on the physical sidelink shared channel (PSSCH), sidelink control information is conveyed through PSCCH/SCI, and sidelink feedback may be conveyed through PSFCH. The gNB may allocate resources and distribute or validate semantic model state. A network-routed alternative, UE$_1\!\rightarrow$gNB$\!\rightarrow$UE$_2$, can be recovered by treating the path as two Uu hops plus relay processing.

\section{Overhead-Aware Analytical Model}
\subsection{Scenario A: Point-to-Point}
For conventional point-to-point communication, the spectral resource cost can be interpreted as
\begin{equation}
A_{\conv}^{\pTwoP}=\frac{L}{\eta_{\conv}}+A_{\conv,\mathrm{hdr}}+A_{\conv,\mathrm{ack}}+A_{\conv,\rs}.
\end{equation}
For SemCom,
\begin{equation}
A_{\sem}^{\pTwoP}=\frac{\rho L}{\eta_{\sem}}+A_{\meta}+A_{\sem,\mathrm{ack}}+A_{\mathrm{align}}+ A_{\sem,\rs}+\frac{A_{\mathrm{model}}}{N},
\end{equation}
where $A_{\conv,\mathrm{hdr}}$ is the conventional header cost, $A_{\conv,\mathrm{ack}}$ and $A_{\sem,\mathrm{ack}}$ are acknowledgement (ACK) feedback, $A_{\meta}$ is semantic metadata, $A_{\mathrm{align}}$ is semantic alignment signaling, $A_{\rs}$ is reference-signal overhead, and $A_{\mathrm{model}}$ is model synchronization overhead.
Similarly, the energy costs are
\begin{align}
E_{\conv}^{\pTwoP}=&e_{\conv}L+E_{\conv,\mathrm{hdr}}+E_{\conv,\mathrm{ack}}+E_{\conv,\rs}+E_{\conv,\mathrm{codec}},\\
E_{\sem}^{\pTwoP}=&e_{\sem}\rho L+E_{\meta}+E_{\sem,\mathrm{ack}}+E_{\mathrm{align}}+E_{\sem,\rs} \nonumber\\
&+E_{\mathrm{enc}}+E_{\mathrm{dec}}+\frac{E_{\mathrm{model}}}{N}.
\end{align}
Thus, point-to-point SemCom is spectrally beneficial only if
\begin{equation}
\frac{\rho L}{\eta_{\sem}}+\Delta A_{\oh}^{\pTwoP}+\frac{A_{\mathrm{model}}}{N}<\frac{L}{\eta_{\conv}},
\label{eq:p2pcondition}
\end{equation}
with
\begin{align}
\Delta A_{\oh}^{\pTwoP}= &A_{\meta}+A_{\sem,\mathrm{ack}}+A_{\mathrm{align}}+A_{\sem,\rs} \nonumber\\
&-A_{\conv,\mathrm{hdr}}-A_{\conv,\mathrm{ack}}-A_{\conv,\rs}.
\end{align}
\vspace{-6mm}
\subsection{Scenario B: UE-to-gNB}
\subsubsection{UE-gNB uplink}
For the conventional uplink from UE to gNB, the spectral cost is expressed as
\begin{align}
A_{\conv}^{\Uu}=&\frac{L}{\eta_{\mathrm{PUSCH},\conv}}+A_{\mathrm{SR}}+A_{\mathrm{grant}}+A_{\mathrm{BSR},\conv} \nonumber\\
&+A_{\mathrm{DMRS}}+A_{\mathrm{HARQ},\conv}+A_{\mathrm{PDCP/RLC}}.
\end{align}
For semantic uplink,
\begin{multline}
A_{\sem}^{\Uu}=\frac{\rho L}{\eta_{\mathrm{PUSCH},\sem}}+A_{\mathrm{SR}}+A_{\mathrm{grant}}+A_{\mathrm{BSR},\sem} 
+A_{\mathrm{DMRS}}\\+A_{\mathrm{HARQ},\sem}+A_{\mathrm{sem\mbox{-}meta}}+A_{\mathrm{sem\mbox{-}fb}} 
+\frac{A_{\mathrm{model},\Uu}}{N}.
\end{multline}
Here, PUSCH is the physical uplink shared channel, SR is the scheduling request, BSR is the buffer status report, PDCP/RLC denotes packet data convergence protocol/radio link control overhead, and $A_{\mathrm{model},\Uu}$ is the Uu model-synchronization resource cost.
The semantic metadata term is decomposed as
\begin{align}
A_{\mathrm{sem\mbox{-}meta}} = &A_{\mathrm{taskID}}+A_{\mathrm{modelID}}+A_{\mathrm{version}}+A_{\mathrm{codebook}} \nonumber\\
&+A_{\mathrm{QoS}}+A_{\mathrm{confidence}}.
\end{align}
The UE-side energy comparison is
\begin{align}
E_{\conv,\mathrm{UE}}^{\Uu}=&E_{\mathrm{PUSCH},\conv}+E_{\mathrm{PUCCH},\conv}+E_{\mathrm{PDCCH\mbox{-}mon}} \nonumber\\
&+E_{\conv,\mathrm{codec}},
\end{align}
\vspace{-6mm}
\begin{align}
E_{\sem,\mathrm{UE}}^{\Uu}=&E_{\mathrm{PUSCH},\sem}+E_{\mathrm{PUCCH},\sem}+E_{\mathrm{PDCCH\mbox{-}mon}}\nonumber \\
&+E_{\mathrm{enc}}+E_{\mathrm{sem\mbox{-}meta}}+\frac{E_{\mathrm{model},\Uu}}{N}.
\end{align}
Here, PUCCH is the physical uplink control channel, PDCCH-mon denotes physical downlink control channel monitoring, and $E_{\mathrm{enc}}$ and $E_{\conv,\mathrm{codec}}$ denote semantic encoder and conventional codec energy, respectively.
The common scheduling terms cancel only if SemCom uses the same scheduling mode and numerology as the conventional baseline. Otherwise, $\Delta A_{\ctrl}^{\Uu}$ and $\Delta E_{\ctrl}^{\Uu}$ must be retained explicitly.
\subsubsection{gNB-to-multi-UE downlink}
For the conventional downlink, denoted by superscript $gd$, the gNB transmits an independent
unicast stream to each UE. Let $K$ denote the number of served UEs. The total spectral-resource consumption is
\vspace{-3mm}
\begin{multline}
A_{\conv}^{gd}
=K\Bigg(
\frac{L}{\eta_{\mathrm{PDSCH},\conv}}
+A_{\mathrm{DCI}}
+A_{\mathrm{DMRS}}
\\+A_{\mathrm{HARQ},\conv}
+A_{\mathrm{PDCP/RLC}}
\Bigg),
\vspace{-3mm}
\end{multline}
where PDSCH is the physical downlink shared channel and $A_{\mathrm{DCI}}$ denotes the spectral cost for downlink control information (DCI). For semantic downlink, a common semantic representation is
transmitted once and reused by multiple UEs with similar semantic
objectives. The corresponding spectral-resource consumption becomes
\vspace{-2mm}
\begin{multline}
A_{\sem}^{gd}= \frac{\rho L}{\eta_{\mathrm{PDSCH},\sem}}
+A_{\mathrm{DCI}}
+A_{\mathrm{DMRS}}
+A_{\mathrm{HARQ},\sem} 
+A_{\mathrm{sem\mbox{-}meta}}\\
+A_{\mathrm{sem\mbox{-}fb}}
+\frac{A_{\mathrm{model},gd}}{N}+K\,A_{\mathrm{UE\mbox{-}specific}},
\end{multline}
where
$A_{\mathrm{UE\mbox{-}specific}}$
denotes the per-UE signaling that cannot be shared,
such as UE-specific feedback, control signaling,
or reliability-related procedures.

The corresponding gNB-side energy consumption is
\begin{equation}
E_{\conv,\mathrm{gNB}}^{gd}
=
K\left(
E_{\mathrm{PDSCH},\conv}
+
E_{\conv, \mathrm{ctrl}}
+
E_{\conv,\mathrm{codec}}\right),
\end{equation}
\vspace{-4mm}
\begin{align}
E_{\sem,\mathrm{gNB}}^{gd}
=&
E_{\mathrm{PDSCH},\sem}
+
E_{\sem,\mathrm{ctrl}}
+
E_{\mathrm{enc}} \nonumber
\\
&
+
E_{\mathrm{sem\mbox{-}meta}}
+
K\,E_{\mathrm{UE\mbox{-}specific}}
+
\frac{E_{\mathrm{model},gd}}{N}.
\end{align}

Unlike the conventional downlink, where the payload transmission
scales linearly with the number of users, semantic communication
transmits a shared semantic representation only once.
Therefore, only the UE-specific signaling and feedback overheads
grow approximately linearly with $K$, while the semantic payload
and model synchronization costs are amortized over multiple users.
\vspace{-1mm}
\subsection{Scenario C: UE-to-UE under One gNB}
\subsubsection{UE-to-UE sidelink}
For gNB-controlled sidelink,
\vspace{-1mm}
\begin{align}
A_{\conv}^{\SL}=&\frac{L}{\eta_{\mathrm{PSSCH},\conv}}+A_{\mathrm{PSCCH},\conv}+A_{\mathrm{SCI},\conv} \nonumber\\
&+A_{\mathrm{PSFCH},\conv}+A_{\mathrm{SL\mbox{-}DMRS}}+A_{\mathrm{Uu\mbox{-}sched},\conv},
\end{align}
where PSSCH is the physical sidelink shared channel, PSCCH carries sidelink control, SCI carries sidelink control information, PSFCH carries sidelink feedback, SL-DMRS denotes sidelink demodulation reference signals, and $A_{\mathrm{Uu\mbox{-}sched}}$ captures gNB scheduling assistance. For semantic sidelink,
\begin{align}
A_{\sem}^{\SL}=&\frac{\rho L}{\eta_{\mathrm{PSSCH},\sem}}+A_{\mathrm{PSCCH},\sem}+A_{\mathrm{SCI},\sem}+A_{\mathrm{PSFCH},\sem} \nonumber\\
&+A_{\mathrm{SL\mbox{-}DMRS}}+A_{\mathrm{Uu\mbox{-}sched},\sem}+A_{\mathrm{sem\mbox{-}meta}}^{\SL} \nonumber\\
&+A_{\mathrm{sem\mbox{-}fb}}^{\SL}+\frac{A_{\mathrm{model},\SL}}{N}.
\end{align}
Here,
\begin{align}
A_{\mathrm{sem\mbox{-}meta}}^{\SL}=&A_{\mathrm{taskID}}+A_{\mathrm{modelID}}+A_{\mathrm{version}}+A_{\mathrm{featureFormat}}\nonumber \\
&+A_{\mathrm{semanticQoS}}+A_{\mathrm{UEcontext}}.
\end{align}
The sidelink energy costs are
\begin{align}
E_{\conv}^{\SL}=&E_{\mathrm{UE1,tx}}^{\conv}+E_{\mathrm{UE2,rx}}^{\conv}+E_{\mathrm{PSCCH},\conv} \nonumber\\
&+E_{\mathrm{PSFCH},\conv}+E_{\mathrm{Uu\mbox{-}sched},\conv}+E_{\conv,\mathrm{codec}},
\end{align}
\vspace{-4mm}
\begin{multline}
E_{\sem}^{\SL}=E_{\mathrm{UE1,tx}}^{\sem}+E_{\mathrm{UE2,rx}}^{\sem}+E_{\mathrm{PSCCH},\sem}+E_{\mathrm{PSFCH},\sem} \\
+E_{\mathrm{Uu\mbox{-}sched},\sem}+E_{\mathrm{enc},\mathrm{UE1}}+E_{\mathrm{dec},\mathrm{UE2}}+\frac{E_{\mathrm{model},\SL}}{N}.
\end{multline}
\subsubsection{UE-to-gNB-to-UE routed}
For the network-routed alternative UE$_1\!\rightarrow$gNB$\!\rightarrow$UE$_2$, the same notation gives
\begin{equation}
A_x^{\mathrm{routed}} = A_{x,\mathrm{UL}}^{\Uu}+A_{x,\mathrm{DL}}^{\Uu}+A_{x,\mathrm{relay}},
\end{equation}
where UL, DL, and relay denote uplink, downlink, and gNB relay-processing costs. An analogous expression applies for energy. This alternative increases both control and receive energy at the gNB, but may reduce direct sidelink coordination requirements.

\section{Break-Even Analysis}
\subsection{Spectral Break-Even}
Define the scenario-dependent overhead difference
\begin{align}
\Delta A_{\oh}^{(s)} =& A_{\sem,\meta}^{(s)} + A_{\sem,\fb}^{(s)}-A_{\conv,\fb}^{(s)} \nonumber\\
&+A_{\sem,\ctrl}^{(s)}-A_{\conv,\ctrl}^{(s)}+A_{\sem,\rs}^{(s)}-A_{\conv,\rs}^{(s)}.
\end{align}
Assuming a single effective payload spectral efficiency for each scenario $s$, SemCom is spectrally beneficial when
\begin{equation}
A_{\sem}^{(s)} < A_{\conv}^{(s)}.
\end{equation}
Substituting \eqref{eq:spectralgeneric} gives the payload-size condition
\begin{equation}
L > L_{\min}^{A,(s)} =
\frac{\Delta A_{\oh}^{(s)}+A_{\sem,\sync}^{(s)}/N}
{1/\eta_{\conv}^{(s)}-\rho/\eta_{\sem}^{(s)}}.
\label{eq:spectral_be}
\end{equation}
Equation \eqref{eq:spectral_be} is meaningful only when
\begin{equation}
\frac{1}{\eta_{\conv}^{(s)}}>\frac{\rho}{\eta_{\sem}^{(s)}}.
\end{equation}
If this condition fails, the semantic payload is not sufficiently compact relative to its spectral efficiency, and no amount of payload scaling can compensate for positive overhead.

\subsection{Energy Break-Even}
Let $e_{\conv}^{(s)}$ and $e_{\sem}^{(s)}$ denote effective communication energy per payload bit for conventional and semantic transmission, respectively. Let
\begin{align}
\Delta E_{\oh}^{(s)}=&E_{\sem,\meta}^{(s)}+E_{\sem,\fb}^{(s)}-E_{\conv,\fb}^{(s)} \nonumber\\
&+E_{\sem,\ctrl}^{(s)}-E_{\conv,\ctrl}^{(s)}+E_{\sem,\rs}^{(s)}-E_{\conv,\rs}^{(s)}.
\end{align}
Then SemCom is energy beneficial if
\begin{equation}
L > L_{\min}^{E,(s)} =
\frac{\Delta E_{\oh}^{(s)}+E_{\sem,\cmp}^{(s)}-E_{\conv,\cmp}^{(s)}+E_{\sem,\sync}^{(s)}/N}
{e_{\conv}^{(s)}-\rho e_{\sem}^{(s)}}.
\label{eq:energy_be}
\end{equation}
The denominator requires
\begin{equation}
 e_{\conv}^{(s)} > \rho e_{\sem}^{(s)}.
\end{equation}
Thus, even if SemCom reduces transmitted payload, it may fail the energy test when neural inference, memory movement, synchronization, or receive-side processing dominates.

\subsection{Joint Efficiency Condition}
For equal utility $U_0$, SemCom is net-efficient in scenario $s$ only if
\begin{equation}
L > L_{\min}^{(s)} = \max\left\{L_{\min}^{A,(s)}, L_{\min}^{E,(s)}\right\}.
\label{eq:joint_be}
\end{equation}
Equation \eqref{eq:joint_be} is the central design rule of this paper. It states that SemCom is not inherently efficient; rather, it becomes efficient only after the semantic payload savings amortize practical overheads. The most important variables are the semantic compression factor $\rho$, synchronization reuse factor $N$, neural compute energy $E_{\sem,\cmp}^{(s)}$, and scenario-specific control overhead $\Delta A_{\oh}^{(s)}$.

\subsection{Design Implications}
The analytical expressions provide several protocol-level implications. First, short-packet SemCom is vulnerable to fixed metadata, feedback, synchronization, and scheduling costs, because $L$ may be smaller than \eqref{eq:joint_be}. Second, stable applications with long model or knowledge-base reuse intervals are more favorable, since synchronization terms scale as $1/N$. Third, UE-centric SemCom must be evaluated using device-side energy, because inference and memory access can dominate transmit-bit savings.

For future 3GPP evolution, e.g., Release 20 and beyond toward 6G, SemCom should be standardized as an overhead-aware protocol capability rather than only an application-layer compression method. This suggests lightweight signaling for semantic task IDs, model or knowledge-base IDs, versioning, representation formats, and semantic QoS targets. Semantic feedback should also be distinguished from conventional ACK/NACK, allowing receivers to report task-level utility, confidence, or semantic failure. Finally, model and knowledge-base synchronization should support cache validity, update periodicity, and multi-UE sharing, so that SemCom gains can be realized without ignoring metadata, feedback, computation, and synchronization costs.

\section{Simulation Results}
The numerical evaluation compares SemCom and conventional communication under equal utility. Unless otherwise stated, $\rho=0.3$, $K=10$, $N=1000$, and $L_0=10^5$ bits. The table columns P2P, Uu, gd, SL, and routed denote point-to-point, UE--gNB Uu, gNB downlink, sidelink, and network-routed UE-to-UE cases. The parameters $\eta_C$ and $\eta_S$ are conventional and semantic spectral efficiencies; $A_{\conv,\mathrm{oh}}$ and $A_{\sem,\mathrm{oh}}$ collect fixed protocol, metadata, feedback, and reference-signal overheads; $A_{\mathrm{sync}}$ is amortized synchronization overhead; $e_C$ and $e_S$ are energy per payload bit; and $E_{\mathrm{oh}}$, $E_{\mathrm{cmp}}$, and $E_{\mathrm{sync}}$ are fixed overhead, computation, and synchronization energy costs. Table \ref{tab:numerical_parameters} summarizes the values.
\begin{table}[!t]
\centering
\scriptsize
\setlength{\tabcolsep}{3pt}
\renewcommand{\arraystretch}{0.92}
\caption{Illustrative parameters for the numerical example.}
\label{tab:numerical_parameters}
\begin{tabular}{lccccc}
\hline
Parameter & P2P & Uu & gd & SL & routed\\
\hline
$\eta_C$ & 2.0 & 1.5 & 2.5 & 1.8 & $1.5+2.5$\\
$\eta_S$ & 1.8 & 1.35 & 2.2 & 1.6 & $1.35+2.2$\\
$A_{\conv,\mathrm{oh}}$ & 200 & 1200 & $500K$ & 1600 & 2600\\
$A_{\sem,\mathrm{oh}}$ & 600 & 2500 & $4000+400K$ & 3600 & 5200\\
$A_{\mathrm{sync}}$ & $10^6$ & $2\times 10^6$ & $4\times10^6$ & $2.5\times10^6$ & $2.5\times10^6$\\
$e_C$ (nJ/bit) & 0.002 & 0.010 & $0.006K$ & 0.004 & 0.010+0.006\\
$e_S$ (nJ/bit) & 0.002 & 0.010 & 0.006 & 0.004 & 0.010+0.006\\
$E_{\conv,\mathrm{oh}}$ (nJ) & 10 & 80 & $25K$ & 120 & 180\\
$E_{\sem,\mathrm{oh}}$ (nJ) & 40 & 160 & $300+20K$ & 240 & 340\\
$E_{C,\mathrm{cmp}}$ (nJ) & 20 & 20 & $10K$ & 30 & 30\\
$E_{S,\mathrm{cmp}}$ (nJ) & 300 & 600 & $1000+400K$ & 900 & 1000\\
$E_{\mathrm{sync}}$ (nJ) & $10^5$ & $2\times10^5$ & $4\times10^5$ & $3\times 10^5$ & $2.5\times10^5$\\
\hline
\end{tabular}
\end{table}

\begin{figure}[!t]
    \centering

    \begin{subfigure}[t]{0.7\linewidth}
        \centering
         \includegraphics[width=1\linewidth]{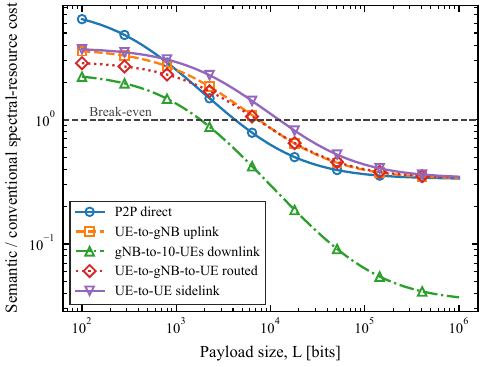}
   \caption{Spectral cost ratio of SemCom over conventional communication.}
        \label{fig:spectral_cost}
    \end{subfigure}
    \begin{subfigure}[t]{0.7\linewidth}
        \centering
        \includegraphics[width=1\linewidth]{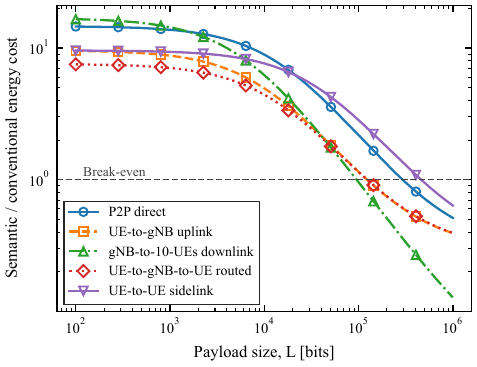}
   \caption{Energy cost ratio of SemCom over conventional communication.}
        \label{fig:energy_cost}
    \end{subfigure}

    \caption{Resource cost ratio of SemCom over conventional communication.}
    \label{fig:payload_size_vs_cost}
\end{figure}

Fig. \ref{fig:spectral_cost} and Fig. \ref{fig:energy_cost} compare the spectral and energy cost ratios between SemCom and conventional communication. Both ratios decrease as payload size increases, because fixed semantic overheads are amortized over more useful information. For short payloads, metadata, feedback, control, and scheduling costs can outweigh semantic compression gains, especially in cellular scenarios. All considered scenarios eventually become spectrally beneficial, while the energy ratio remains above one over a wider range due to semantic encoding/decoding, metadata handling, and model synchronization energy.

\begin{figure}[!t]
    \centering

    \begin{subfigure}[t]{0.7\linewidth}
        \centering
         \includegraphics[width=1\linewidth]{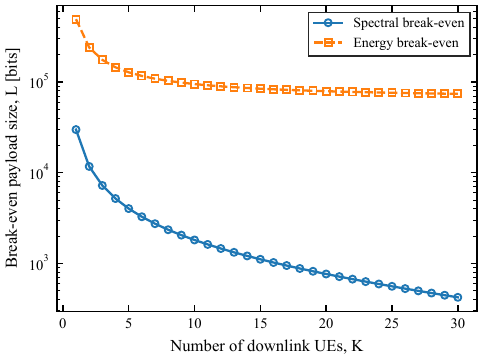}
   \caption{Multi-user downlink amortization gain under different numbers of UEs.}
        \label{fig:multi_UE1}
    \end{subfigure}
    \begin{subfigure}[t]{0.7\linewidth}
        \centering
        \includegraphics[width=1\linewidth]{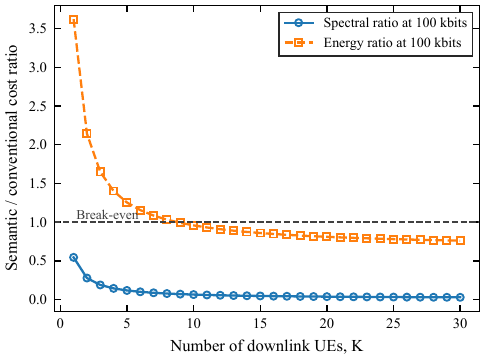}
   \caption{Multi-user downlink cost ratio versus number of UEs.}
        \label{fig:multi_UE2}
    \end{subfigure}

    \caption{Break-even payload size and semantic/conventional cost ratio under different numbers of UEs.}
    \label{fig:multi_UE}
\end{figure}

Fig. \ref{fig:multi_UE1} and Fig. \ref{fig:multi_UE2} show the amortization gain of multi-user downlink SemCom. As $K$ increases, the spectral break-even payload decreases rapidly because the common semantic representation and shared overheads are reused across more UEs, while the energy threshold decreases more gradually due to persistent encoder, synchronization, and UE processing costs. At $L_0=10^5$ bits, the spectral ratio remains well below one and the energy ratio eventually crosses the break-even threshold, confirming that multi-user downlink is particularly favorable for SemCom.

\begin{figure}[!t]
    \centering

    \begin{subfigure}[t]{0.7\linewidth}
        \centering
         \includegraphics[width=1\linewidth]{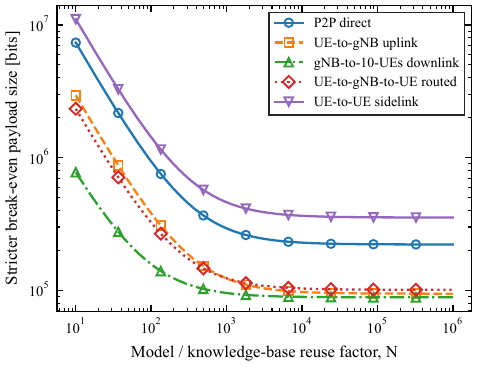}
   \caption{Reuse sensitivity of overhead-aware SemCom.}
        \label{fig:reuse_factor}
    \end{subfigure}

    \begin{subfigure}[t]{0.7\linewidth}
        \centering
        \includegraphics[width=1\linewidth]{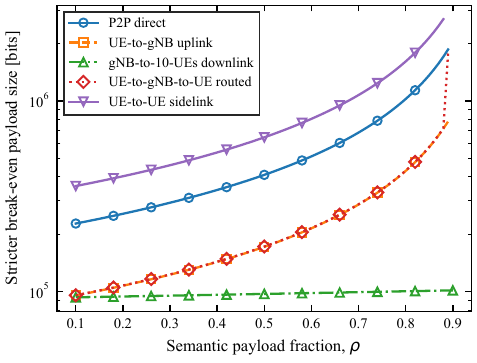}
   \caption{Payload size sensitivity under different semantic compression factor $\rho$.}
        \label{fig:compression_ratio}
    \end{subfigure}

    \caption{Break-even payload size versus reuse factor $N$ and semantic compression factor $\rho$.}
    \label{fig:payload_size_vs_N_rho}
\end{figure}

Fig. \ref{fig:payload_size_vs_N_rho} further investigates the effect of the reuse factor $N$ and the semantic compression ratio $\rho$ on the break-even payload size. Fig. \ref{fig:reuse_factor} illustrates the impact of the model reuse factor on the stricter break-even payload size. For all considered scenarios, increasing the reuse factor $N$ significantly reduces the break-even payload size, especially when the reuse factor is small. As $N$ increases, the curves gradually converge, indicating that the synchronization cost becomes negligible and the remaining break-even payload size is dominated by transmission cost and fixed protocol overhead. Moreover, Fig. \ref{fig:compression_ratio} shows the sensitivity of the break-even payload size to the semantic payload fraction $\rho$. As $\rho$ increases, the transmission advantage of SemCom is weakened compared with conventional communication, leading to a larger break-even payload size. These results indicate that both model or knowledge-base reuse and semantic compression are key factors that determine the practicality of SemCom.

\section{Conclusion}
This paper quantified practical SemCom overhead using a unified spectral-resource and energy model across point-to-point, NR Uu, multi-user downlink, sidelink, and routed UE-to-UE deployments. The break-even conditions show that SemCom is beneficial only when semantic compression and model reuse amortize protocol, synchronization, and computation overheads. Numerical results confirm that short-payload SemCom is often inefficient, energy gains are stricter than spectral gains, and multi-user downlink, large model reuse, and strong compression are key enablers.



\section*{Acknowledgment}
This work is supported by the 6G-GOALS project under the 6G SNS-JU Horizon program, n.101139232.
\vspace{-1mm}

\bibliographystyle{IEEEtran} %
\bibliography{references} 
\end{document}